\documentclass{an}
\usepackage{graphicx}
\usepackage{times}
\usepackage{fancyhdr}
\begin{document}

\title{Theoretical approaches to particle propagation and acceleration in
turbulent intergalactic medium}

\author{A. Lazarian}
\institute{Dept. of Astronomy, University of Wisconsin-Madison, USA}

\date{Received; accepted; published online}

\abstract{Intercluster medium is expected to be turbulent with turbulence being superAlfvenic
at large scales. Magnetic fields substantially modify the turbulent cascade when the turbulence
reaches the scales at which the fluctuation velocity gets less than the Alfven velocity. At those scales
it is possible to consider three cascades, of fast, slow and Alfven modes with little energy exchange
between them. As Alfvenic and slow modes are anisotropic they marginally scatter and accelerate
cosmic rays, while fast modes dominate the processes. However, in the presence of cosmic rays the
 turbulence is modified as cosmic rays transfer the energy of compressible motions (i.e. slow and
fast modes) from large  scales to the scale of cosmic ray Larmor radius. This
results in generation of a new small-scale  Alfvenic component which is not a part of the ordinary 
MHD cascade. This component does scatter and accelerate cosmic rays. In addition, magnetic 
reconnection in turbulent medium accelerates cosmic rays. The complexity of the interacluster 
turbulence calls for observational studies. A new technique Velocity Coordinate Spectrum (VCS) is
particularly promising for studies of velocity fluctuations with a new generation of X-ray observatories. 
 }

\correspondence{lazarian@astro.wisc.edu}

\maketitle

\section{Do we expect IGM to be turbulent?}

A fluid of viscosity $\nu$ becomes turbulent when the rate of viscous dissipation, which is  $\sim \nu/L^2$ at the energy injection scale $L$, is much smaller than the energy transfer rate $\sim V_L/L$, where $V_L$ is the velocity dispersion at the scale $L$. The ratio of the two rates is the Reynolds number $Re=V_LL/\nu$. In general, when $Re$ is larger than $Re_{crit}\sim 
10-100$ the system becomes turbulent. Chaotic structures develop gradually as  $Re$ increases, and those with $Re\sim10^3$ are appreciably less chaotic than those with $Re\sim10^8$.

It is widely accepted that medium in clusters should be turbulent.
Mergers of galactic subclusters may be one of the major energy injection
mechanism (see Sarazin 2002, Brunetti 2003 and references therein). The details
of the injection and the energy transfer are poorly understood.
A crude picture includes the energy injection scale of 100-500 kpc and
the injection velocity of the order of $10^3$ km/s. 

A difficulty that one faces trying to understand turbulence is clusters
is that the diffusivity is very different along and perpendicular to
magnetic field. This is related to the fact that the mean free path of
ions is substantial. For {\it non-magnetized} intracluster gas the
$Re$ number is only marginally larger than $Re_{crit}$ and
therefore formally is only just sufficient for initiating the turbulence.
At the same time, even small magnetization makes the Reynolds number for
the motions perpendicular to magnetic field very large (e.g. more than
$10^{10}$). This poses the question of what a reasonable choice of
the Reynolds number is for intracluster medium.

Strangely enough we do not have a good answer to this basic question. There
are several processes that definitely influence the effective diffusivity
of intracluster plasma. First of all, compressions of collisionless
plasma should result
in various instabilities that would decrease the mean free paths of protons.
In addition, bending of magnetic field lines does not allow 
mean free path of protons to be larger than the scale at which the turbulent
velocity is equal to the Alfven velocity (henceforth, the $l_A$ scale). 
Therefore the effective Reynolds number at the injection scale will be
$\sim 10^3$ for the field of 1 $\mu$G field, which is the field
that fills according to Enslin et al. (2005) 90\% of the intracluster volume.

Initially the turbulence is superAlfvenic and hydrodynamic motions easily bend
magnetic field lines. Such turbulence is analogous to hydrodynamic one until
the scales of the order of $l_A\equiv L (V_A/V_L)^3$ are reached. 
Assuming $V_L=10^3$ km/s and $L=100$ kpc we get 30~pc, if we adopt, following Ensslin et al (2005)
that 1 $\mu$G field fills 90\% of the ingracluster volume. The 
turbulence gets mangetohydrodynamic (MHD) for scales $l<l_A$ as magnetic fields control fluid 
motions. However, in both ``hydrodynamic'' and MHD regimes 
plasma effects continue to be important as they drain and redistribute energy
of compressible motions (see Schekochihin et al. 2005).

\section{What do we know about MHD turbulence?}

There have long been understanding that the MHD turbulence
is anisotropic (e.g. Shebalin et al.~1983). Substantial progress has been achieved
by Goldreich \& Sridhar (1995; hereafter GS95), who made an
ingenious prediction regarding relative motions parallel and
perpendicular to magnetic field {\bf B} for incompressible
MHD turbulence. 
An important observation that leads to understanding of the GS95
scaling is that magnetic field 
cannot prevent mixing motions
of magnetic field lines if the motions
are perpendicular to the magnetic field. Those motions will cause, however,
waves that will propagate along magnetic field lines.
If that is the case, 
the time scale of the wave-like motions along the field, 
i.e. $\sim l_{\|}/V_A$,
($l_{\|}$ is the characteristic size of the perturbation along 
the magnetic field and 
$V_A=B/\sqrt{4 \pi \rho}$ is 
the local Alfven speed) will be equal to the hydrodynamic time-scale, 
$l_{\perp}/v_l$, 
where $l_{\perp}$ is the characteristic size of the perturbation
perpendicular to the magnetic field.
The mixing motions are 
hydrodynamic-like.
They obey Kolmogorov scaling,
$v_l\propto l_{\perp}^{1/3}$,  because incompressible turbulence is assumed.
Combining the two relations above
we can get the GS95 anisotropy, $l_{\|}\propto l_{\perp}^{2/3}$
(or $k_{\|}\propto k_{\perp}^{2/3}$ in terms of wave-numbers).
If  we interpret $l_{\|}$ as the eddy size in the direction of the 
local 
magnetic field.
and $l_{\perp}$ as that in the perpendicular directions,
the relation implies that smaller eddies are more elongated.
The latter is natural as it the energy in hydrodynamic motions
decreases with the decrease of the scale. As the result it gets more
and more difficult for feeble hydrodynamic motions to bend magnetic 
field lines. 

GS95 predictions have been confirmed 
numerically (Cho \& Vishniac 2000; Maron \& Goldreich 2001;
Cho, Lazarian \& Vishniac 2002, 2003); 
they are in good agreement with observed and inferred astrophysical spectra 
(see Cho \& Lazarian 2005). What happens in a compressible MHD? Does any part
of GS95 model survives? The answer depends on the mode coupling.
According to closure calculations (Bertoglio, 
Bataille, \& Marion 2001; see also Zank \& Matthaeus 1993),
the energy in compressible modes in {\it hydrodynamic} turbulence scales
as $\sim M_s^2$ if $M_s<1$.
Cho \& Lazarian (henceforth CL03) conjectured that this relation can be extended to MHD turbulence
if, instead of $M_s^2$, we use
$\sim (\delta V)_{A}^2/(a^2+V_A^2)$. 
(Hereinafter, we define $V_A\equiv B_0/\sqrt{4\pi\rho}$, where
$B_0$ is the mean magnetic field strength.) 
However, since the Alfven modes 
are anisotropic, 
this formula may require an additional factor.
The compressible modes are generated inside the so-called
Goldreich-Sridhar cone, which takes up $\sim (\delta V)_A/ V_A$ of
the wave vector space. The ratio of compressible to Alfvenic energy 
inside this cone is the ratio given above. 
If the generated fast modes become
isotropic (see below), the diffusion or, ``isotropization'' of the
fast wave energy in the wave vector space increase their energy by
a factor of $\sim V_A/(\delta V)_A$. This  results in
\begin{equation}
\frac{\delta E_{comp}}{\delta E_{Alf}}\approx \frac{\delta V_A V_A}{V_A^2+a^2},
\label{eq_high2}
\end{equation}
which suggests that the drain of energy from
Alfvenic modes is marginal along the cascade.

Our considerations above about the mode coupling can guide us
in the discussion below. Indeed,
if Alfven cascade evolves on its own, it is natural to assume that 
slow modes exhibit the GS95 scaling.
Indeed, slow modes in gas 
pressure dominated environment (high $\beta$ plasmas) are
similar to the pseudo-Alfven modes in incompressible regime 
(see GS95; Lithwick \& Goldreich 2001). The latter modes do follow
the GS95 scaling. 
In magnetic pressure dominated environments  or low $\beta$ plasmas, 
slow modes are density perturbations propagating with the
sound speed $a$ parallel to the mean magnetic field. 
Those perturbations are essentially
static for $a\ll V_A$. 
Therefore Alfvenic turbulence is expected to mix density
perturbations as if they were passive scalar. This also induces the
GS95 spectrum.

The fast waves in low $\beta$ regime propagate at $V_A$ irrespectively
of the magnetic field direction. 
In high $\beta$ regime, the properties of fast modes are similar, 
but the propagation speed is the sound speed $a$.
Thus the mixing motions induced by Alfven waves should marginally
affect the fast wave
cascade. It is expected to
be analogous to the acoustic wave cascade and hence be isotropic.

\section{How does turbulence scatter and accelerate cosmic rays?}

The propagation of cosmic rays (CRs) is affected by their interaction
with magnetic field. This field is turbulent and therefore, the resonant
interaction of cosmic rays with MHD turbulence has been discussed
by many authors as the principal mechanism to scatter and isotropize
cosmic rays. Although cosmic ray diffusion can
happen while cosmic rays follow wandering magnetic fields (Jokipii 1966), the acceleration of 
cosmic rays requires efficient scattering.
For instance, scattering of cosmic rays back into the shock is a
vital component of the first order Fermi acceleration.

While most investigations are restricted to Alfv\'{e}n modes propagating
along an external magnetic field (the so-called slab model of Alfv\'{e}nic
turbulence) (see Schlickeiser 2002), obliquely propagating MHD modes have been included in
Fisk et al. (1974) and later studies (see Pryadko \& Petrosian  1999). A more complex models
 were obtained by combining the 
results of the Reduced MHD with parallel slab-like
modes . Models that better correspond to the current understanding of MHD turbulence
(see above) have been considered lately. Chandran (2000, henceforth C00) considered resonant 
scattering and acceleration by incompressible MHD turbulence. Resonant scattering and
acceleration by the compressible MHD turbulence was considered in Yan \& Lazarian (2002, 2004,
henceforth YL02, YL04).
There the following result was obtained for the diffusion coefficients that governed by
 Alfvenic scattering
\begin{eqnarray}
\left[\begin{array}{c}
D_{\mu\mu}\\
D_{pp}\end{array}\right]&=&\frac{v^{2.5}\mu^{5.5}}{\Omega^{1.5}L^{2.5}(1-\mu^2)^{0.5}}\Gamma[6.5,k_{max}^{-\frac{2}{3}}k_{\parallel,res}L^{\frac{1}{3}}]\nonumber\\
& &\left[\begin{array}{c}
1\\
m^{2}V_{A}^{2}\end{array}\right],\label{ana}
\end{eqnarray}
where $\Gamma[a,z]$ is the incomplete gamma function. This result was obtained using
a tensor of magnetic fluctuations that was obtained in the CLV02 study. It provides 
scattering and acceleration rates orders of
magnitude larger than those
 in C00 for the most of energies considered. However, if energy is injected through random driving
at large
scale $L\gg k_{\parallel,res}^{-1}$ 
the scattering frequency,
\begin{equation}
\nu=2D_{\mu\mu}/(1-\mu^{2}),\label{nu}
\end{equation}
are still much smaller
than the estimates for isotropic and slab model. 
What does
scatter cosmic rays? Our work in YL02 identified fast modes as the principal agent responsible
for CR scattering and acceleration. A study in YL04 showed that this is true in spite of the fact
that fast modes are subjected to much more dissipation compared to the Alfven modes. In the
next section we shall discussed another possibility proposed in Lazarian \& Beresnyak (2006), namely, instabilities related to cosmic rays that inject Alfven modes at resonant scales (see \S 4). This
possibility may provide a physical justification for the earlier calculations in Brunetti et al. (2004). 

The work above was done in relation to galactic cosmic rays. However, the problem of MHD
turbulence with CR is a general one. Therefore in a recent paper Cassano \&
Brunetti (2005) considered fast modes as the principal component of CR acceleration in galaxy
clusters (see also Brunetti 2004). A further work in this direction is in Brunetti \& Lazarian (2006).

Resonant acceleration (we include Transit-Time-Damping acceleration to this category) is not the 
only process by which magnetized turbulence accelerates CR. For instance, the
acceleration of cosmic rays by the large scale compressible motions was described in the literature
rather long time ago (see Ptuskin 1988). In this regime slow and fast diffusion limit exist. The slow
limit corresponds to the rate of particle diffusion out of compressible eddies, which is slower than the 
evolution rate of the eddies. On the contrary, in the fast diffusion limit particles leave the eddies before they turnover. Large scale compressions associated with
anisotropic slow modes were used to accelerate Solar Flare CR in Chandran (2003) who considered
slow diffusion regime.
Fast diffusion regime was used in Chandran \& Maron (2004ab) in application to CR acceleration in
galaxy clusters. A comprehensive study of these processes has been done in Cho \& Lazarian (2006).
We found, first of all, in slow diffusion limit the resonance scattering, that is  a part and parcel of
the slow diffusion process, is the dominant acceleration process. Then, we identified fast modes as the principal cause of non-resonant acceleration. In addition, we showed that weak turbulence (Galtier
et al. 2005) may be
important for cosmic ray acceleration in the fast diffusion limit.  

While the importance of fast  modes for the scattering and  the  acceleration of CR is unquestionable,
plasma and CR instabilities may make the actual turbulence in galaxy clusters more  involved compared to a simple picture presented above. For instance, below we consider a robust CR instability that should modify the properties of MHD turbulence.

\section{How can cosmic rays modify turbulence?}

Studies of the backreaction of the energetic particles on the
turbulence are usually limited by the damping of turbulence on
energetic particles (see Brunetti \& Blasi 2005,
Ptuskin et al. 2005, Petrosian, Yan \& Lazarian 2006). However,
this is not the only effect of CR. For instance, Lazarian \& Beresnyak
(2006) have considered a transfer of turbulent energy to small (but
not dissipation!) scales
that is mediated by CR.  As CR present a collisionless fluid,
 compressions CR through the compression of magnetic field
 preserve the  adiabatic invariant  $p^2_\perp/B$, where $p_\perp$ is the CR momentum
perpendicular to magnetic field.  CR with the anisotropic
distribution of momenta, i.e. with nonzero 
$A=(p_\perp-p_\|)/p_\|$ are subjected to an instability, which growth rate can be estimated as
\begin{equation}
\Gamma_i(k_{\|}) = \omega_{pi} \frac{n_{CR}(p>m\Omega/k_{\|})}{n}AQ,
\label{stream}
\end{equation}
where $\Omega=eB/mc$ is a cyclotron frequency, $m$ -- proton mass, $n$
is the density of plasma, $\omega_{pi}$ is the ion plasma frequency,
refering to $n$ and $n_{CR}(p>m\Omega/k)$ is the number density of
cosmic rays with momentum larger than minimal resonant momentum for a
wavevector value of $k$. $Q$ is a dimensionless numerical factor, depending
only on cosmic ray power-law index $\alpha$.
\begin{figure}
\includegraphics[width=.5\textwidth]{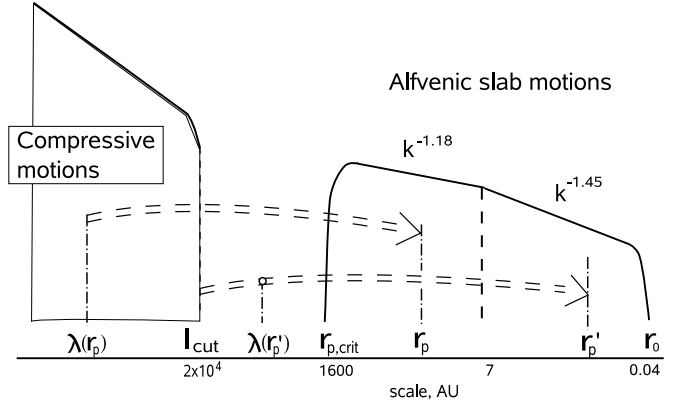}
  \caption{Energy density of compressive modes and Alfv\'enic slab-type
waves, induced by CRs, in galaxy clusters. The energy is transferred from
the mean free path scale to the CR Larmor radius scale. If mean free path
falls below compressive motions cutoff, spectrum of slab waves becomes
steeper (from Lazarian \& Beresnyak, 2006).}
\end{figure}

The outcome of this instability is the direct transfer of the energy from mean free path of
a CR, which is determined by magnetic scattering of the CR, to the CR gyroradius. The magnetic
perturbations at the gyroradius scale are the most efficient in CR scattering and this decreases
the mean free path $\lambda$ and therefore $A\sim v_{\lambda}/V_A$.
At large scales the instability is limited
by damping arising from the ambient Alfvenic turbulence ($r_{p, crit}$ in Fig.~1), while at small 
scales steepening limits the amplitude the perturbations. All in all, the interaction of CR with MHD 
turbulence results in an additional slab-like  small-scale component of Alfvenic perturbations that
 interacts with CR much more efficiently than the Alfvenic mode being excited by the energy injection 
scale.

\section{Can magnetic reconnection accelerate CR?}

Magnetic reconnection is a process that we expect to happen routinely
 in magnetized
turbulent fluid. This even more true for superAlfvenic turbulence
 when fluid motions bend magnetic fields easily. Indeed, magnetic
reconnection should happen whenever magnetic field with non-parallel
direction interact.

Several schemes of magnetic reconnection are known (see Fig.~2). 
The most famous one is
the Sweet-Parker (Parker 1957, Sweet 1958) 
reconnection, which naturally occurs when magnetic fields
get into contact over long current sheets. This is the most natural arangement
of magnetic field to get into within the turbulent fluid as random moving
flux tubes collide and push their way through one another. The scheme provide
ridiculously low rates of reconnection to be of astrophysical importance,
however. Indeed, the reconnection rate scales as $V_A Rm^{-1/2}$, where
$Rm=V_A L/\eta_{mag}$ is the Lunquist number, which for realistic values
of magnetic diffusivity $\eta_{mag}$ is so humangous that any processing
of magnetic energy via reconnection gets absolutely negligible for any
astrophysical system (e.g. stars, interstellar medium) not to speak about 
intracluster medium with its much larger scales.

The Petcheck (1964) scheme emerged as an answer to the evidence 
of fast reconnection that cannot be explained within the Sweet-Parker 
model. Within this scheme the reconnection is concentrated along
thin filaments from which oppositely directed magnetic field lines diverge.
It is clear
at this moment that it cannot be sustained  at large $R_L$ 
for smooth resistivities.
Whether or not anomalous effects, e.g. those related to Hall term, 
can provide reconnection at rates comparable with the Alfv\'en speed
is hotly debated\footnote{ For instance, the necessary condition
for the anomalous effects to be important, e.g. for
that the electron mean free path is less than the current sheet thickness (see 
Trintchouk et al. 2003) is difficult to satisfy for the ISM, where
the Sweet-Parker current sheet thickness is typically {\it much} larger
than the ion Larmor radius.} (see Biskamp, Schwarz \& Drake 1997, 
Shay et al. 1998, Shay \& Drake 1998,
Bhattacharjee, Ma \& Wang 2001). We feel, however, that 
the issue of
satisfying boundary outflow conditions is  the most controversial element
in applying Petscheck scheme to astrophysical conditions. If very special
global geometry or magnetic fluxes, e.g. convex magnetized regions, 
is required to enable reconnection, then
for a generic astrophysical case the reconnection is slow (see Fig. 3).

\begin{figure}
\includegraphics[width=.45\textwidth]{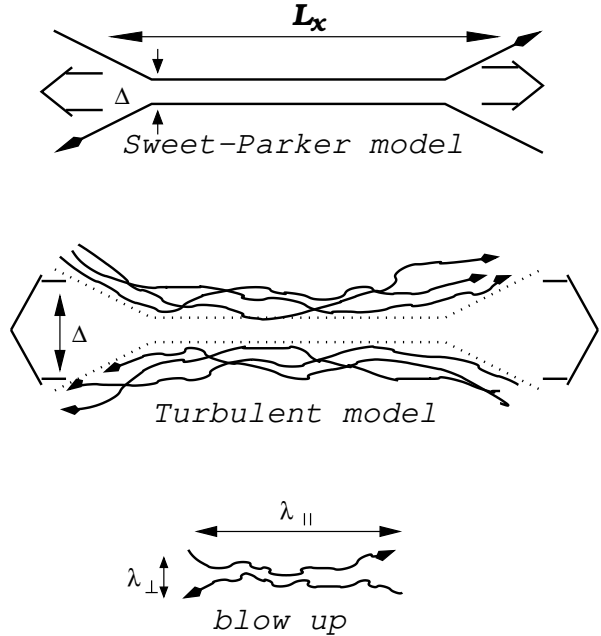}
\caption{ {\it Upper plot}: 
Sweet-Parker model of reconnection. The outflow
is limited by a thin slot $\Delta$, which is determined by Ohmic 
diffusivity. The other scale is an astrophysical scale $L\gg \Delta$.
{\it Middle plot}: Turbulent reconnection model that accounts for the 
stochasticity
of magnetic field lines. The outflow is limited by the diffusion of
magnetic field lines, which depends on field line stochasticity.
{\it Low plot}: An individual small scale reconnection region. The
reconnection over small patches of magnetic field determines the local
reconnection rate. The global reconnection rate is substantially larger
as many independent patches come together.}
\end{figure}

The turbulent reconnection model
proposed in Lazarian \& Vishniac (1999, henceforth LV99) deals with
magnetic field configurations with flat long current sheets. 
However, these sheets consist of small sheets related to the interaction
of individual turbulent elements of magnetic flux. The outflow within
the model is limited not by the thickness of the current sheet, but by
magnetic field wondering. As the result the rate of reconnection is 
proportional to the turbulence intensity. The model can explain flaring
associated with reconnection as well as high rates of reconnection that are
required by both observations and theory (e.g. dynamo theory). Calculations
in Lazarian, Vishniac \& Cho (2004) confirm main points of the LV99 model
(e.g. the rate of field line wondering) 
and extend it to reconnection in partially ionized gas.

\begin{figure}
\includegraphics[width=0.45\textwidth]{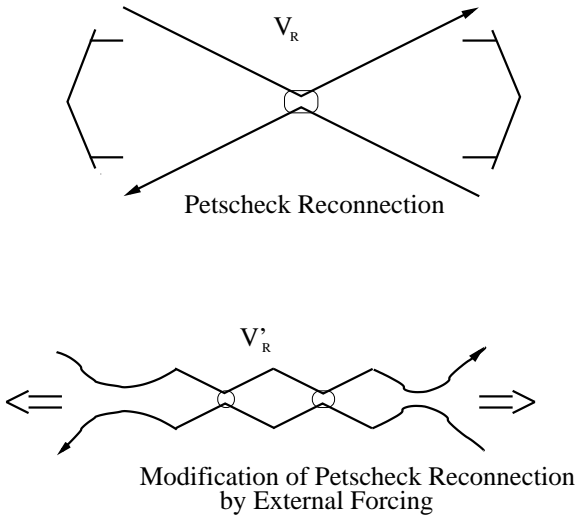}
\caption{ {\it Upper plot}:
Petscheck reconnection scheme has a magnetic diffusion region (rectangular
area at the tip of the magnetic field line bending) for which both
the  longitudinal and transversal dimensions are determined 
by the Ohmic diffusivity.
To enable $V_R\sim V_A$ the model requires field line opening over the whole astrophysical scale
involved, which is difficult to satisfy in practice. {\it Lower plot}:
External forcing, e.g. the forcing present in the ISM, is likely
to close the opening required by the Petscheck model. In this case
the global outflow constraint is not satisfied and the resulting
reconnection speed is $V_R'\ll V_A$. As the result the Petscheck reconnection
cannot operate steadily and therefore cannot deal with the amount of flux
that, for instance, an astrophysical dynamo would require to reconnect.}
\end{figure}

 There are several ways how magnetic 
reconnection can accelerate cosmic rays. It is well known
that electric fields in the current sheet can do the job. For Sweet-Parker
reconnection this may be an important process in those exceptional
instances when Sweet-Parker reconnection is fast in
astrophysical settings. For Petscheck reconnection only an insubstantial
part of magnetic energy is being released within the reconnection
zone, while bulk of the energy is being released in shocks that
support  X-point. Therefore one would expect the shock acceleration
of cosmic rays to accompany Petscheck reconnection. 

\begin{figure}
\includegraphics[width=.35\textwidth]{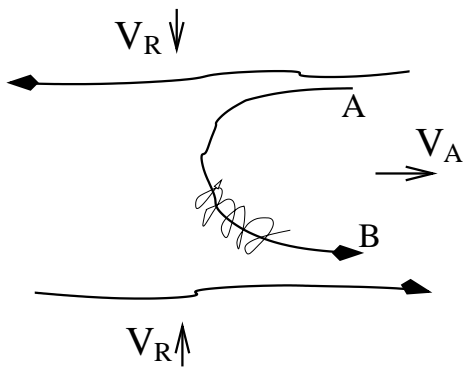}
\caption{Cosmic rays spiral about a reconnected magnetic
field line and bounce back at points A and B. The reconnected
regions move towards each other with the reconnection velocity
$V_R$. The advection of cosmic rays entrained on magnetic field
lines happens at the outflow velocity, which is in most cases
of the order of $V_A$. Bouncing at points A and B happens
because either of streaming instability or turbulence in the
reconnection region.}
\end{figure}

Similarly to the Petscheck scheme 
the turbulent reconnection process assumes that only
small segments of magnetic field lines enter the reconnection zone
and are subjected to Ohmic annihilation. Thus only small fraction of
magnetic energy, proportional to ${R}_L^{-2/5}$ (LV99), is
released in the current sheets. The rest of the energy is
released in the form of non-linear Alfv\'en
waves that are generated as reconnected magnetic field lines straighten
up. Such waves are likely to cause second order Fermi acceleration.
This idea was briefly discussed in Lazarian et al. (2001) in relation
to particle acceleration during the gamma-ray burst events. In addition,
large amplitude Alfv\'enic motions in low $\beta$, i.e. magnetically
dominated, plasmas are likely to induce shocks (see Beresnyak, Lazarian
\& Cho 2005), which can also cause particle acceleration.

However, the most interesting process is the first-order Fermi acceleration
that is intrinsic to the turbulent reconnection. To understand it
consider a particle entrained on a reconnected 
magnetic field line (see Fig.~4). This particle
may bounce back and force between magnetic mirrors formed by oppositely
directed magnetic fluxes moving towards each other with the velocity
$V_R$. Each of such bouncing will increase the energy of a particle
in a way consistent with the requirements of the first-order Fermi
process. The interesting property of this mechanism that potentially
can be used  to test observationally the idea is that the 
resulting spectrum is different from those arising from shocks.
Gouveia Dal Pino \& Lazarian (2003) used particle acceleration
within turbulent reconnection regions to explain
the synchrotron power-law spectrum arising from the flares of the
microquasar GRS 1915+105. Note, that the mechanism acts in the
Sweet-Parker scheme as well as in the scheme of turbulent reconnection.
However, in the former the rates of reconnection and therefore the
efficiency of acceleration are marginal in most cases.

\section{How can we test the turbulence model?}

There are several ways that information about turbulence in intracluster
medium can be obtained from observations. For instance, magnetic power
spectrum has been obtained using Faraday rotation (see Ensslin 2004, Ensslin, Vogt
\& Pfromer 2005). The issues of detectibility of velocity turbulence
have been discussed in Sunyaev, Norman \& Bryan  (2003). Recent advances
in the techniques of studies of turbulence via  velocity fluctuations (see review by
Lazarian 2004 and references therein) enable us to get spectra of turbulent fluctuations.
The techniques described there can potentially separate compressible and incompressible
motions.
For instance, a Velocity Coordinate Spectrum (VCS) technique can be used to
extract turbulence statistics from the Doppler-shifted spectra even when the
measurements are made over a limited number of directions (even one) or
the object is not resolved (Lazarian \& Pogosyan 2006). Such measurements
would already be possible, if not for the failure of the high velocity resolution
instrument on board of the ASTRO-E2. An example of such
a study that can be available with the forthcoming X-ray telescope 
Constellation X is presented in Fig.~5. 

\begin{figure}
\includegraphics[width=.45\textwidth]{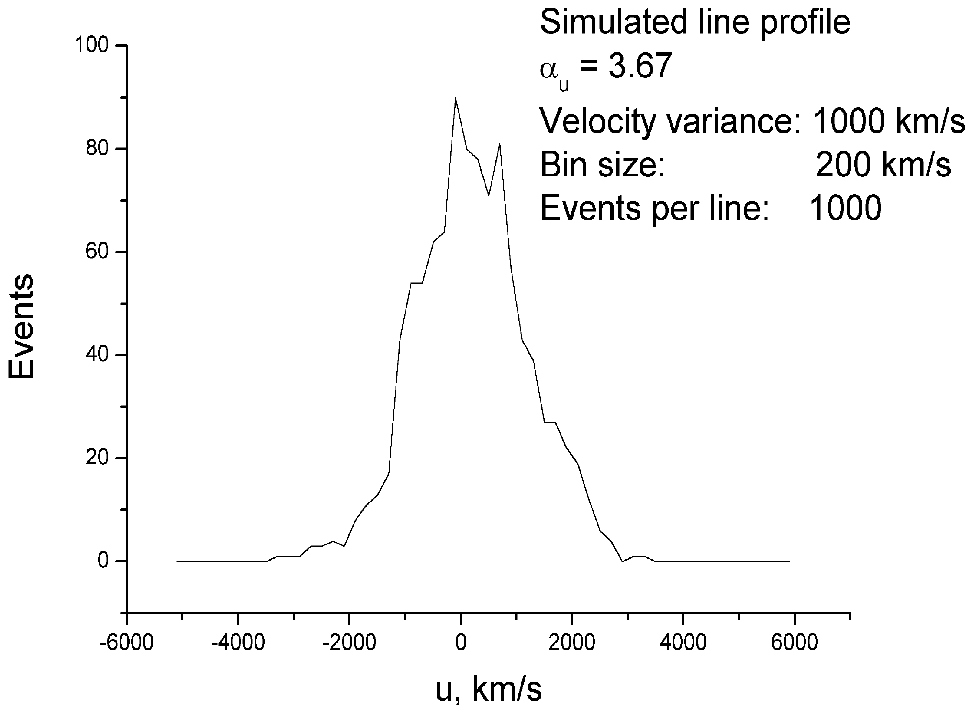}
\includegraphics[width=.45\textwidth]{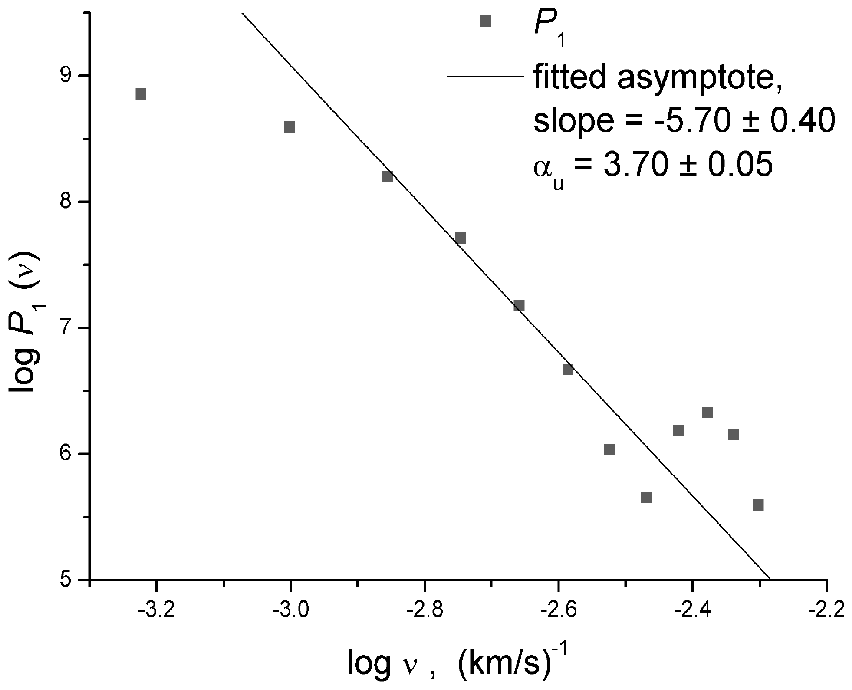}
\caption{An example of the velocity line profile (upper plot) and velocity spectrum (lower plot) for underlying 
Kolmogorov turbulence that can be produced with observations using Constellation X forthcoming
X-ray facilty with 1 hour exposure (Chepurnov \& Lazarian 2006).}
\end{figure}

\section{What are the prospects of the field?}

The detailed description of the CR acceleration and scattering is a challenging
problem that requires coordinated efforts of both theorists and observers.
It is clear that at the moment we do not have an adequate description of 
MHD turbulence in the collisionless fluid and this is an impediment for the
progress. To get such a description one needs to account for different
instabilities that affect  magnetized collisionless intercluster
plasmas. This, together with the change of effective $Re$ fluid number as
magnetic field and instabilities evolve, make the description of turbulence
quite challenging. Together with turbulence driving 
the above processes determine the efficiency of the CR interaction with
magnetized turbulence. We know by now that if the energy is injected at
large scales, the interaction of CR of low energies with Alfvenic part of
MHD cascade is suppressed. This, however, may not be true if some part of 
Alfven modes is generated by small scale plasma instabilities. In particular,
some of these instabilities may be due to CR themselves. In addition, 
reconnection within chaotic fields may accelerate CR directly.
Therefore, with more observational data, clusters of galaxies 
Nevertheless, with  more observational data, clusters of galaxies may serve
 as a good testing ground for studies of the fundamental processes involving CR.



\acknowledgements
 AL acknowledges 
the support from the NSF Center for Magnetic Self-Organization in
Laboratory and Astrophysical Plasmas.



\begin{thebibliography}{}
\bibitem{} Beresnyak, A., Lazarian, A., Cho, J. 2005, ApJ 624, L93
\bibitem{} Bertoglio, J.P., Bataille, F., Marion, J.D. 2001, Phys. Fluids 13, 290
\bibitem{} Biskamp, D., Schwarz, E., Drake, J.F. 1997, Phys. Plasmas 4, 1002
\bibitem{} Bhattacharjee, A., Ma, Z.W., Wang 2001, Phys. Plasmas 8, 1829
\bibitem{} Brunetti, G. in: eds. Bowyer, S., Hwang, C.Y., {\it Matter and Energy in Clusters of
 Galaxies}, 
ASP Vonf. Ser. V. 301, San Fransisco, 349
\bibitem{} Brunetti, G. 2004, JKAS, 37, 583
\bibitem{} Brunetti, G., Blasi, P., Cassano, R., Gabici, S. 2004, MNRAS 350, 1174
\bibitem{} Brunetti, G., Blasi, P. 2005, MNRAS 363, 1173
\bibitem{} Brunetti, G., Lazarian, A. 2006, MNRAS, in prep. 

\bibitem{} Cassano, R., Brunetti, G. 2005, MNRAS 357, 1313

\bibitem{} Chandran, B. 2000, Phys. Rev. Lett., 85 4656
\bibitem{} Chandran, B. 2003, ApJ, 599, 1426 (C03)
\bibitem{} Chandran, B., Maron, J. 2004a, ApJ 602, 170
\bibitem{} Chandran, B., Maron, J. 2004b, ApJ 603, 23

\bibitem{} Cho, J., Lazarian, A. 2002, Phys. Rev. Lett.  24, 5001
\bibitem{} Cho, J., Lazarian, A. 2003, MNRAS, 345 325 (CL03)
\bibitem{} Cho, J., Lazarian, A. 2005, Theo. Comp. Fluid Mech. 19, 127
\bibitem{} Cho, J., Lazarian, A., Vishniac, E. 2002, ApJ 564, 291
\bibitem{} Cho, J., Lazarian, A., Vishniac, E. 2003,
              in: eds. E. Falgarone \&  T. Passot, {\it Turbulence and magnetic fields in astrophysics}, 
                 Lect.~Notes~Phys. Vol.~614,
      Berlin: Springer, p. 56 
\bibitem{} Cho, J., Vishniac, E. 2000, ApJ 539, 273 

\bibitem{} Ensslin, T.A. 2004, JKAS 37, 439
\bibitem{} Ensslin, T.A., Vogt, C., Pfrommer C. 2005, in: eds. K.T. Chyzy, K. Otminowska-Mazur, M. Soida
and R.-J. Dettmar, {\it Magnetized 
Plasma in Galaxy Evolution},  Krakow, p. 231

\bibitem{} Fisk, L. A., Goldstein, M. L., Sandri, G. 1974, ApJ 190, 417 

\bibitem{} Galtier S., Nazarenko S.V., Newell A.C., Pouquet A., 2000,
J. Plasma Phys. {\bf 63}(5), 447-488

\bibitem{} Goldreich P., Sridhar S., 1995, ApJ 438, 763 (GS95)

\bibitem{} Jokipii, J. R. 1966, ApJ 146, 480
 
\bibitem{} Lazarian, A. 2004, JKAS, 37, 563
\bibitem{} Lazarian, A., Beresnyak, A. 2006, MNRAS, submitted, astro-ph/0606737
\bibitem{} Lazarian, A., Petrosian, V., Yan, H. Cho, J. 2002, in 
Beaming and Jets in Gamma Ray Bursts, ed. R. Ouyed (Stanford: NBSI), 45
\bibitem{} Lazarian, A., Pogosyan, D. 2006, ApJ, in press, astro-ph/0511248
\bibitem{} Lazarian A., Vishniac, E.T. 1999, ApJ 517, 700
\bibitem{} Lazarian, A., Vishniac, E., Cho, J.  2004, ApJ 603, 180 

\bibitem{} Lithwick, Y., Goldreich, P. 2001, ApJ 562, 279

\bibitem{} Maron, J., Goldreich, P. 2001, ApJ 554, 1175

\bibitem{} Parker, E.N. 1957, J. Geophys. Res., 62, 509

\bibitem{} Petschek, H.E. 1964, in: ed. W.H. Hess, {\it The Physics of Solar Flares}, AAS-NASA
Symposium, NASA SP-50, Greenbelt, Maryland, p.~425

\bibitem{} Petrosian, V., Yan, H., Lazarian, A. 2005, ApJ, in press

\bibitem{} Pryadko, J.M., Petrosian, V. 1999, ApJ 515, 873 

\bibitem{} Ptuskin, V. 1988, Soviet Astron. Lett. 14, 255 (P88)

\bibitem{} Ptuskin, V.S., Moskalenko, I.V., Jones, F.C., Strong, A.W., Zirakashvili, V.N.
2005, ApJ, sumbitted, astro-ph/0510335


\bibitem{} Sarazin, C.L. 2002, in: eds. Feretti, L.,
Gioia, I.M., Giovannini, G., V., {\it Merging Processes in Clusters of Galaxies}, 272, Kluwer, Dordrecht, 1

\bibitem{} Shay, M.A., Drake, J.F. 1998, Geophys. Res. Lett., 25, 3759
\bibitem{} Shay, M.A., Drake, J.F.,
Denton, R.E., \& Biskamp, D.\ 1998, J. Geophys. Res., 103, 9165

\bibitem{} Shebalin, J. V., Matthaeus, W., Montgomery, D. 
                  1983, J. Plasma Phys. 29, 525

\bibitem{} Schekochihin, A., Cowley, S., Kulsrud, R., Hammett, G. Sharma, P. 2005,
in:  eds. K. Chyzy, K. Otminowska-Mazur, M. Soida, 
and R.-J. Dettmar, {\it Magnetised Plasma in Galaxy Evolution}, Krakow, 86

\bibitem{} Schlickeiser, R. 2002, Cosmic ray astrophysics, Berlin: Springer

\bibitem{} Sunyaev, R.A., Norman, M.L., Bryan, G.L. 2003, Astron. Lett. 29, 783

\bibitem{} Sweet, P.A. 1958, in: ed. B. Lehnert,  IAU Symp. 6, {\it Electromagnetic 
Phenomena in 
Cosmical Plasma},  New York: Cambridge Univ. Press, 123

\bibitem{} Yan, H., Lazarian, A. 2002, Phys. Rev. Lett. 89, 281102 (YL02)

\bibitem{} Yan, H., Lazarian, A. 2004, ApJ 614, 757 (YL04)

\bibitem{} Zank, G.P., Matthaeus, W. H. 1993, Phys. Fluids A 5(1), 257

\end{thebibliography}
\end{document}